\documentclass[twocolumn,pre,aps,showpacs,groupedaddress]{revtex4-1}

\def\alphaloc{{\alpha_{\rm loc}}}

\usepackage{amsmath,amssymb}
\usepackage{pstricks}
\usepackage{epsfig}
\usepackage[latin1]{inputenc}

\psset{unit=1mm}

\def\<{\left<}
\def\>{\right>}
\def\ket|#1>{\left|#1\right>}
\def\bra<#1|{\left<#1\right|}
\def\elem<#1|#2|#3>{\left<#1\right|#2\left|#3\right>}
\def\({\left(}
\def\){\right)}

\def\pl{\partial}

\begin{document}

\title{Intrinsic geometry approach to surface kinetic roughening}

\author{Javier Rodr\'{\i}guez-Laguna$^{1,3}$}\email{jrlaguna@math.uc3m.es}
\author{Silvia N.\ Santalla$^2$}
\author{Rodolfo Cuerno$^1$}
\affiliation{$^1$Mathematics Department and Grupo Interdisciplinar de
Sistemas Complejos (GISC), Universidad Carlos III de Madrid, Leganés (Madrid), Spain \\
$^2$Physics Department, Universidad Carlos III de Madrid, Leganés (Madrid), Spain\\
$^3$ICFO--Institute of Photonic Sciences, Castelldefels (Barcelona), Spain}

\date{April 1, 2011}

\begin{abstract}

A model for kinetic roughening of one-dimensional interfaces is
presented within an intrinsic geometry framework that is free from the
standard small-slope and no-overhang approximations. The model is
meant to probe the consequences of the latter on the
Kardar-Parisi-Zhang (KPZ) description of non-conserved, irreversible
growth. Thus, growth always occurs along the local normal direction to
the interface, with a rate that is subject to fluctuations and depends
on the local curvature. Adaptive numerical techniques have been
designed that are specially suited to the study of fractal
morphologies and can support interfaces with large slopes and
overhangs. Interface self-intersections are detected, and the ensuing
cavities removed. After appropriate generalization of observables such
as the global and local surface roughness functions, the interface
scaling is seen in our simulations to be of the Family-Vicsek type for
arbitrary curvature dependence of the growth rate, KPZ scaling
appearing for large sytems sizes and sufficiently large noise
amplitudes.
\end{abstract}


\maketitle

%%%%%%%%%%%%%%%%%%%%%%%%%%%%%%%%%%%%%%%%%%%%%%%%%%%%%%%%%%%%%

\section{Introduction}

Kinetic roughening in non-equilibrium interfacial growth has been
investigated for a long time both as a fundamental problem in
statistical mechanics and as a model to relevant physical systems,
such as dynamics of thin films, fluid flow, flame front propagation,
or undifferentiated biological growth \cite{Barabasi}. Starting from a
flat shape, for many interfaces it is generally observed that they
roughen until eventually a steady state is attained, in which the
interface roughness, $W$ (root-mean-square deviation around the mean
interface position) scales in a nontrivial way with the system size
$L$, as $W\sim L^\alpha$, where $\alpha$ is the so-called roughness
exponent. At short times in these systems, the roughness scales as a
power law of time, $W(t)\sim t^\beta$, with $\beta$ being called the
growth exponent \cite{Barabasi}. The saturation roughness can be
measured at smaller length scales $l$ than the system size, another
power law ensuing, $W(l)\sim l^\alphaloc$. If the local and global
roughness exponents are equal, $\alphaloc=\alpha$, the system follows
the so-called Family-Vicsek (FV) Ansatz, which is a generalization of
that expected for equilibrium critical dynamics \cite{Barabasi};
otherwise, {\em anomalous scaling} occurs
\cite{Krug_AP97,Ramasco_PRL00} in which interface fluctuations follow
different laws at small and large scales, a fact that is not uncommon
when the interface profile displays many large jumps and high slope
values \cite{Asikainen_PRE02,Asikainen_EPJB02}. In general, the values
of the critical exponents $\alpha$, $\alphaloc$, and $\beta$ identify
universality classes that play an important role in the classification
of nonequilibrium processes in spatially extended systems
\cite{Odor_RMP04}.

Fluctuations arise in the above physical systems due to a variety of
sources, such as non-homogeneous driving, randomness of growth events,
etc. Dynamically, and in the absence of specific conservation laws
that constrain the evolution of the interface \cite{Krug_AP97}, they
compete with irreversible growth, which can be simply represented by a
constant growth rate along the local normal direction to the
interface, and are attenuated through various relaxation
processes. The latter can be differential absorption due to curvature
effects (e.g., condensation of atoms from a vapor phase is more
effective at surface valleys of a solid than at surface peaks)
\cite{Kardar_PRL86}, to surface diffusion (atoms move from peaks to
valleys) \cite{Villain_JP91,Lai_PRL91,Pimpinelli}, etc. Different
relaxation mechanisms can give rise to different universality classes,
some of which may in general appear as transients during the system
evolution.

One of the most fruitful approaches to surface kinetic roughening is
the study of stochastic partial differential equations (SPDE) of a
certain {\em height function}, $h(\mathbf{x},t)$, that represents the
surface height position at time $t$ above point $\mathbf{x}$ on a
reference plane, in the so-called {\em Monge gauge}
\cite{Safran,Marsili_RMP96}. This implies the assumption that the
interface has no {\em overhangs}. Moreover, most height equations in
this context rely on a {\em small-slope} approximation that allows to
strongly simplify the analytical formulation. However, this
simplification may be at the cost of inaccuracies in the physical
description. For instance, surface diffusion is thus taken along the
horizontal substrate plane (instead of along the interface) and is
severely overestimated in regions of high slopes, giving rise to large
{\em grooves} as in the so-called linear Molecular Beam Epitaxy (MBE)
equation \cite{Family_Fra93}.

%A natural question arises as to what extent
%scaling properties can change in the absence of both the no-overhang
%and the small-slope approximations.

The very same Kardar-Parisi-Zhang (KPZ) equation \cite{Kardar_PRL86},
that is one of the most important SPDEs within the study of kinetic
roughening ---and within non-equilibrium statistical mechanics
indeed--- was actually formulated within both the no-overhang and the
small slope approximations. The fact that the corresponding roughness
exponent $\alpha_{\rm KPZ} < 1$ \cite{Barabasi} makes this
approximation self-consistent, in the sense that slopes remain small
on average. Nevertheless, the derivation of this equation being based
on rather general considerations, a natural question arises as to what
is the effect of relaxing these approximations. Actually, early
reports \cite{Maritan_PRL92} already claimed that relaxing the
no-overhang approximation in a KPZ-type equation leads to a larger
value for the growth exponent than that characteristic of the standard
KPZ equation, although no information was provided on the value of the
roughness exponent, nor are later confirmations of this result
available.  One of the motivations of the present work is to address a
more systematic study along these lines. The question becomes all the
more interesting since experimental realizations of KPZ scaling are
remarkably scarce, in spite of the wide asymptotic universality that
it is expected to have on theoretical grounds
\cite{Cuerno_04,Cuerno_EPJB07}.  Nevertheless, this universality class
has been indeed verified in experiments very recently
\cite{Takeuchi_PRL10}, including detailed predictions
\cite{Sasamoto_PRL10,Amir_CPAM11} on the asymptotic height
distribution for the case of growth in a circular geometry. Analogous
predictions are also available for the simpler case of a {\em band
  geometry} \cite{Praehofer_PRL00} in which boundary conditions are
periodic or free, and the initial interface is a straight, horizontal
line. Still, these remain to be experimentally assessed
\cite{Takeuchi_PRL10}.

Actually, the relation between the scaling properties of a given SPDE
in band and in circular geometries has been the focus of a recent
controversy that has arisen from conflicting analytical predictions
\cite{Escudero_PRL08,Krug_PRL09,Escudero_PRL09b}. These unavoidably
require approximations, since even equations that are linear in a band
geometry become nonlinear in the circular setting. Thus, in order to
address the issue, it would be interesting to be able to study
numerically the same SPDE in both cases. The circular geometry
requires naturally a multiple valued interface (as seen from a Monge
gauge) and arbitrary interface orientations, the additional existence
of overhangs being again the most complete situation to be
considered. This constitutes a second motivation for the present work,
and makes the formulation of such an equation in a band geometry ---as
in the seminal formulation by KPZ--- a natural preliminary step. Note,
moreover, that the observables (correlation functions, etc.) also need
to be adapted to the type of geometry in which the continuum model is
formulated, see e.g.\ \cite{Bru_MCM08} and references therein. Again
from this point of view, starting out with the band geometry seems a
natural choice.

Previous attempts have been made in order to study continuum models of
kinetic roughening beyond the two approximations mentioned. For
instance, Siegert and Plischke \cite{Siegert_PRL92}, Maritan {\em et
  al}.\ \cite{Maritan_PRL92} or Family and Amar \cite{Family_Fra93}
considered various relaxation operators still in the Monge gauge, but
beyond the small slopes approximation, thus studying very complex and
highly nonlinear SPDEs that are hard to deal with even
numerically. Actually, later developments have questioned the
reliability of the schemes employed, since the standard finite
difference discretization of nonlinear terms has been shown to lead to
numerical inaccuracies and instabilities (see an overview in
e.g.\ \cite{Miranda_PRE08}).  Approximations in the corresponding
SPDEs are also usually performed in the study of continuum growth
equations in circular geometry, typically including the neglect of
overhangs as well as additional simplifications, see
e.g.\ \cite{Bru_MCM08,Escudero_JSTAT09} and references therein. The
specific case of the KPZ equation has been addressed in some of these
works \cite{Maritan_PRL92,Bru_MCM08} and elsewhere, see
e.g.\ \cite{Kapral_PRE94,Batchelor_PA98}, although overhangs seem to
have been consistently neglected in all these SPDE-based approaches.

Also, phase-field studies have been carried out since the pioneering
work of Keblinski {\em et al}.\ \cite{Keblinski_PRE96,Nicoli_JSTAT09}.
However, while in this class of studies arbitrary interface geometries
are available, computational expense is sufficiently high that, to our
knowledge, no complete systematic study of scaling properties is
available so far. Moreover, connection with the KPZ equation is at
best indirect via the asymptotic behavior, if appropriate.

In this work, our approach is to study a one-dimensional interface as
a 2D curve in a band geometry as described above, that evolves under a
continuum equation whose terms are determined by its intrinsic
geometry. Actually, these are the same as those that led to the
formulation of the KPZ equation, but now free from the no-overhang and
the small-slope approximations. We study the ensuing model
numerically, carrying out simulations with an adaptive number of
points, and doing measurements in a geometrically natural way. The
equations do not distinguish between the horizontal (substrate) and
vertical (growth) directions, their difference residing only in the
initial and boundary conditions.

The paper is organized as follows. Section \ref{model} introduces the
physical model we study and its geometric interpretation. Next, in
Section \ref{numerics}, we discuss the numerical implementation of the
simulations. In Section \ref{measurements} the observables to be
measured are considered, while results are presented in Section
\ref{results}. Further discussion of these is presented in the final
Section \ref{conclusions}, together with some conclusions and
indications of potential new directions for future work.

%%%%%%%%%%%%%%%%%%%%%%%%%%%%%%%%%%%%%%%%%%%%%%%%%%%%%%%%%%%%%

\section{Intrinsic geometry for surface growth in 1D}
\label{model}

\subsection{Intrinsic Geometric Preliminaries}

Our 1D interface is given mathematically by a curve $\gamma$ in a 2D
band geometry, that we take to be an infinite cylinder of
circumference $L$, i.e., $\mathbf{r}: [0,1] \mapsto S^1\times
\mathbb{R}$. Of course, $\mathbf{r}(0)=\mathbf{r}(1)$. The curve is assumed to
wrap around the cylinder exactly once, thus separating a {\em solid
  phase} (below) from a {\em dilute phase} (above).

At any point on the curve $\mathbf{r}$, let us consider the unitary tangent
vector field, $\mathbf{u}_t$, and the normal vector field pointing towards
the dilute phase, $\mathbf{u}_n$. The curvature $K(\mathbf{r})$ {\em along} the
interface will be given by the inverse of the radius of the best
fitting (osculating) circle.

For any scalar field defined on the curve $\phi(\mathbf{r})$, we define its
$\gamma$-gradient as the derivative {\em along} the curve,
parametrized by its arc-length $s$, pointing along the tangent:
\begin{equation}
\nabla_\gamma \phi(\mathbf{r}(s)) = {\pl\phi(\mathbf{r}(s))\over\pl s} \,
\mathbf{u}_t(\mathbf{r}(s)).
\label{gamma.derivative}
\end{equation}
The {\em divergence} along the curve of such vector field allows us to
define a Laplace-Beltrami operator as
\begin{equation}
\nabla^2_\gamma \phi(\mathbf{r}(s)) = {\pl^2\phi(\mathbf{r}(s))\over \pl s^2}.
\label{laplacian}
\end{equation}

\subsection{Physical Model}

In order to have a physical image in mind, let us consider a system in
which atoms are interchanged between a dilute (vapor or liquid) and a
solid (aggregated) phase, in such a way that the latter grows at the
expense of the former, as in the production of thin films
\cite{Cuerno_04}. We will consider absorption and desorption to be
non-directional phenomena, and the possibility of atomic diffusion at
the interface. The chemical potential difference at an interface
point, $\mu(\mathbf{r})$, will be the only driving force for growth. On the
solid side of the interface (subindex $s$), the chemical potential
will be given by \cite{Krug_AP97}
\begin{equation}
\mu_s(\mathbf{r})=\mu_{s,0} - \mu_{s,1}K(\mathbf{r}),
\label{chem.pot.surf}
\end{equation}
\noindent where $\mu_{s,0}$ and $\mu_{s,1}>0$ are empirical
parameters, and the local curvature $K(\mathbf{r})$ provides a measure of
the number of local ``dangling bonds''. We will consider $K$ to be
positive when the interface is concave as seen from the dilute
phase. On the other hand, the chemical potential in the dilute
(``vapor", subindex $v$) phase is $\mu_v(\mathbf{r})=
\mu_{v,0}$. Given that the solid phase is growing at the expense of
the dilute phase, we can take $\mu_{s,0} <0$ and $\mu_{v,0}=0$ without
loss of generality \cite{Krug_AP97}.  On the other hand, particles may
move along the surface in their attempt to minimize their chemical
potential. This process is called {\em surface diffusion}
\cite{Mullins_JAP57,Mullins_JAP59}. Locally, the particle current will
be proportional to the {\em surface gradient} of $\mu_s$,
\begin{equation}
\mathbf{j}_s(\mathbf{r})\propto -\nabla_\gamma \mu_s(\mathbf{r}).
\label{diffusion.current}
\end{equation}
\noindent Conservation of the number of particles requires that they
aggregate to the solid phase at a rate proportional to the divergence
of this current \cite{Krug_AP97}, thus
\begin{equation}
\nabla_\gamma \cdot \mathbf{j}_s(\mathbf{r})\propto -\nabla^2_\gamma \mu_s(\mathbf{r}) =
\mu_{s,1} \nabla^2_\gamma K(\mathbf{r}).
\end{equation}

In line with the seminal approach by KPZ \cite{Kardar_PRL86}, a point
at the interface $\mathbf{r}$ is assumed to move along the local normal
direction, with a velocity that is proportional to the rate at which
particles accumulate there, namely,
\begin{equation}
\pl_t \mathbf{r} = \( -\Gamma (\mu_s(\mathbf{r})-\mu_v) -
\nabla_\gamma \cdot \mathbf{j}_s(\mathbf{r}) + \eta_v(\mathbf{r},t) \) \mathbf{u}_n,
\label{vn_vs_mu}
\end{equation}
\noindent where $\Gamma>0$ is a mobility constant \cite{Krug_AP97} and
$\eta_v$ is a zero-average noise term ---describing, e.g.,
fluctuations in the rate of growth events--- that we will take as
Gaussian, with zero average, and white both in space and time,
\begin{equation}
\<\eta_v(\mathbf{r},t)\eta_v(\mathbf{r}',t')\>\propto \delta(\mathbf{r}-\mathbf{r}')\delta(t-t').
\label{noise}
\end{equation}
\noindent Note that the delta function in \eqref{noise} is evaluated
at points {\em on} the interface. Thus, it does not correspond to a
delta function on the horizontal substrate plane.

The general form of the ensuing evolution equation is thus
\begin{equation}
\pl_t \mathbf{r} = \( A_0 + A_1 K(\mathbf{r}) - A_2
\nabla^2_\gamma K(\mathbf{r}) + A_{n} \eta(\mathbf{r},t) \) \mathbf{u}_n.
\label{growth.eq.complete}
\end{equation}
\noindent This equation contains four terms:

\begin{itemize}

\item{} A constant term, $A_0$, measuring the average difference
  between the chemical potentials in the dilute and solid phases.
\item{} A term proportional to the curvature, with coefficient
  $A_1>0$, having the physical interpretation of surface tension
  \cite{Krug_AP97}.
\item{} A term proportional to the Laplacian of the curvature field
  along the curve, with coefficient $A_2>0$.
\item{} A white noise term, due to the fluctuations in growth events,
  with variance $A_n^2$.
\end{itemize}

In this work we study the scaling properties of
Eq.\ \eqref{growth.eq.complete} for $A_2=0$, leaving the general case
for further work. Therefore, the equation under study will be
\begin{equation}
\pl_t \mathbf{r} = \( A_0 + A_1 K(\mathbf{r}) + A_n \eta(\mathbf{r},t)
\) \mathbf{u}_n.
\label{growth.eq}
\end{equation}
\noindent This SPDE is analogous to the KPZ equation in the sense that
it incorporates the same basic ingredients, namely, growth along the
local normal direction, relaxation by surface tension, and
fluctuations, but there are several important differences: (a) growth
is always normal to the interface, while in the KPZ equation this
condition holds only approximately; (b) we employ the full curvature,
which is nonlinear if written in the Monge gauge, and (c) noise is
normal to the interface, which means that, as seen in the Monge gauge,
it is {\em multiplicative} and {\em correlated}.

In order to represent an interface, the curve should be {\em simple}
for all times, i.e.: no multiple points are allowed. Therefore,
Eq.\ (\ref{growth.eq}) should be {\em supplemented} with a condition
regarding the treatment of self-intersections. Our criterion is simply
{\em removal}, keeping the component which wraps around the cylinder.

Finally, an initial condition is required. Unless otherwise specified,
the interface will start as a horizontal line.

%%%%%%%%%%%%%%%%%%%%%%%%%%%%%%%%%%%%%%%%%%%%%%%%%%%%%%%%%%%%%%%%%

\section{Numerical Simulations}
\label{numerics}

The numerical discretization of Eq.\ (\ref{growth.eq}) has been
carried out keeping in mind the desideratum of {\em geometric
  naturality}, using concepts borrowed from {\em discrete differential
  form} theory \cite{Desbrun_05}. The idea of our numerical procedure
is to employ an adaptive scheme that does not particularly privilege
any system direction, and that avoids e.g.\ standard explicit finite
differences discretization of the various differential operators
appearing, since this requires an explicit parametrization of the
interface and is prone in this context to numerical instabilities, see
e.g.\ \cite{Gallego_PRE07} and references therein.  The guiding
physical principle is to use discrete counterparts of the operators
appearing in Eq.\ \eqref{growth.eq}, which are geometrically
motivated.

The interface is simulated numerically by a chain of points
$\{P_1,\ldots,P_N\}$, with adaptive $N$ and periodic boundary
conditions, $P_{N+1}=P_1$. It is implemented in the computer in the
form of a {\em linked list}, ensuring easy insertion and deletion of
points. This set fulfills the following {\em continuity
  condition}\ \cite{Alvarez_PRL98}: the distance between any couple of
neighboring points should always be in the interval
$[l_{min},l_{max}]$. Let us use the symbol $d_{i,j}$ for the distance
between any two points of the set. Then, at every time step, this
condition is ensured using the following algorithm for every couple of
nearest neighbors $P_i$, $P_{i+1}$:
\begin{itemize}
\item{}If $d_{i,i+1}<l_{min}$, remove $P_{i+1}$ from the
  list.
\item{}If $d_{i,i+1}>l_{max}$, add a new point
  $P_{i+1}=(P_i+P_{i+1})/2$ (and shift the indices appropriately).
\end{itemize}
Application of this algorithm ensures that the representation of the
interface is homogeneous. It is important, in all simulations, to
generate the initial points randomly, in order to avoid lattice
artifacts.

Next we provide a discrete approximation to the different geometric
fields needed to simulate Eq.\ (\ref{growth.eq}). The tangent vector
field, $\mathbf{u}_t(x)$ is approximated by $(\mathbf{u}_t)_{i,i+1}\approx
(P_{i+1}-P_i)/d_{i,i+1}$. Notice that these values are not attached to
single points, but to {\em links} between two consecutive points. With
it we can approximate the normal vector field, $(\mathbf{u}_n)_{i,i+1}$,
finding the orthogonal vector to $(\mathbf{u}_t)_{i,i+1}$, pointing towards
the dilute phase.

The curvature field, $K(\mathbf{r})$, is approximated using the
triangle $P_{i-1},P_i,P_{i+1}$. We define $K(P_i)$ as the inverse of
the radius of the circumscribed circle passing through all three
points. Let $\alpha_i$ be the angle at $P_i$. Then,
\begin{equation}
K_i\equiv K(P_i)\approx \frac{2\sin(\alpha_i)}{d_{i-1,i+1}}.
\label{curvature}
\end{equation}
\noindent Notice that, as defined, the sign of the curvature depends
on the sign of $\alpha_i$. As in the previous Section, it will be
positive when the interface is concave, as seen from the dilute phase.

As our last step, we will determine the approximation to the gradient
and Laplacian along the curve of any scalar field $\phi(\mathbf{r})$,
using the notation $\phi_i\equiv\phi(P_i)$,
\begin{equation}
\(\nabla_\gamma \phi\)_{i,i+1} \approx
\frac{\phi_{i+1}-\phi_i}{d_{i,i+1}} (\mathbf{u}_t)_{i,i+1}.
\label{gradient}
\end{equation}
\noindent These values are attached to links, not to points. The
approximation to the Laplace-Beltrami operator is computed as the
(numerical) divergence of that expression:
\begin{equation}
\(\nabla^2_\gamma \phi\)_i \approx \frac{2}{d_{i-1,i}+d_{i,i+1}}
\(\frac{ \phi_{i+1}-\phi_i }{d_{i,i+1}} -
\frac{ \phi_i - \phi_{i-1} }{d_{i-1,i}} \).
\label{laplacian_disc}
\end{equation}
\noindent This formula may be formally justified using the notion of
Hodge dual \cite{Desbrun_05}.

As pointed out before, the interface may tend to create {\em
self-intersections}, leading to cavity formation. Our code detects
such self-intersections and removes one component, keeping the one
which wraps around the cylinder. The detection algorithm is one of the
most delicate and costly ingredients of our numerical scheme.

Finally, concerning the time evolution, the SPDE \eqref{growth.eq} is
simulated using the usual Euler-Maruyama algorithm. We have ensured
that our $\Delta t$ is small enough for the computations by repeating
a selection of them with smaller values, always obtaining convergence.
Regarding the noise amplitude, it is important to remark that, in the
discretized SPDE, it must be multiplied by $(\Delta s)^{-1/2}$, where
$\Delta s$ is the arc-length interval corresponding to each point.
This guarantees that the noise amplitude does not depend on the local
interface orientation \cite{Marsili_RMP96}.

Additional checks have been performed on our algorithm in order to
reproduce expected features in the behavior of Eq.\ (\ref{growth.eq})
(not shown): (1) In the deterministic case ($A_n=0$) with $A_0=0$, we
have performed a linear stability analysis of Eq.\ \eqref{growth.eq},
plotting the relaxation time of a sinusoidal initial condition with
wave-vector $k$, obtaining the expected diffusion-like relation
between typical relaxation time and spatial frequency, $\tau \sim
k^{-2}$. In this sense, the deterministic case of
Eq.\ \eqref{growth.eq} reproduces (non-linear) diffusion. (2) For
small non-zero $A_0$, with $A_n=A_1=0$, a sinusoidal perturbation
evolves towards the typical deterministic KPZ arcade.

%%%%%%%%%%%%%%%%%%%%%%%%%%%%%%%%%%%%%%%%%%%%%%%%%%%%%%%%%%%%%%%%%%%

\section{Measurements}
\label{measurements}

The standard basic observable in kinetic roughening theory is the
global interface {\em width} or surface roughness. It is usually
defined within the Monge gauge as
\begin{equation}
W_{\rm Monge}=\< \overline{\(h- \overline{h}\)^2} \>^{1/2} ,
\label{width.monge}
\end{equation}
\noindent where $h$ is the height of the interface, as measured from a
horizontal reference line, bar stands for space average across the
full system and brackets denote noise realizations. This definition,
therefore, privileges the horizontal orientation.
%, the only reason to do that being the initial condition, which might be
%completely forgotten in the steady state.
%Thus, a tilted flat surface will have a very large width whose
%meaning may be spurious.
Moreover, if the interface presents {\em overhangs},
the very use of $h(x)$ becomes ambiguous.

In many applications and in experimental studies of e.g.\ thin film
dynamics \cite{Zhao}, it is customary to {\em remove the plane}, i.e.,
finding the optimal orientation which can be taken as the
horizontal. Our definition of interface width goes beyond this, being
completely {\em intrinsic} at every length scale. Given any portion of
interface, proceed in the following way:
\begin{itemize}
\item{}Find the straight line which makes the best fit to the curve.
\item{}Compute the root mean square distance of the interface points
to this line.
\end{itemize}
\noindent Skipping the first step and considering the best fitting
line to be horizontal amounts to using the Monge gauge definition of
interface width. Note that the fit is {\em not} an usual least-squares
fit.  In such a case, the minimized distances are taken along vertical
lines. In our case, we minimize the actual distances, i.e., along a
{\em diagonal} line, from the curve points to the fitting line.

Likewise, we define local width $w(l,t)$ at any length scale $l$ in
the usual way: sample the surface using {\em windows} of size $l$, and
measure the average width obtained, following the previous
procedure. Of course, it is expected that $w$ should be an increasing
function of $l$. At the maximum length-scale, $l=L$, we retrieve the
global width $W(t)=w(L,t)$. We should remark that the windows should
be {\em random}. Otherwise, lattice artifacts are bound to appear.

The Monge-gauge roughness at small scales, $\lim_{l\to 0} w_{\rm
  Monge}(l,t)$, describes the {\em slopes statistics}. In our case, if
the surface is smooth below a certain cutoff length, such limit will
be close to zero. If the interface is defined, as in our numerical
simulations, by a set of linked line segments whose lengths lie in a
certain interval, there is an upper bound for the roughness at each
scale.

A power-law behavior is expected, $w\sim l^\alphaloc$, for length
scales under a certain correlation length, $l<\xi(t)$. Length scales
above this value are still uncorrelated. As time evolves, $\xi(t)$
increases until it reaches the system size. At that time all length
scales are correlated, the system saturates to a stationary state and
its width does not increase any further. The correlation length grows
with another power law, $\xi(t)\sim t^{1/z}$, defining the {\em
  dynamic exponent}, $z$. Using this definition, it is easy to check
that the {\em saturation time} scales as $t_s\sim L^z$
\cite{Barabasi}.  Scaling is {\em regular}, or FV type, when the
roughness of the correlated scales stays fixed for all times. In that
case, $\alphaloc=\alpha$ and it can be shown that $z=\alpha/\beta$
\cite{Barabasi}. On the other hand, in the presence of anomalous
scaling, $\alphaloc \neq \alpha$ and the roughness of the correlated
scales keeps increasing after correlation, only saturating when
$\xi(t)=L$ \cite{Krug_AP97,Ramasco_PRL00}.

%%%%%%%%%%%%%%%%%%%%%%%%%%%%%%%%%%%%%%%%%%%%%%%%%%%%%%%%%%%%%%%%%%%

\section{Results}
\label{results}

Our main result is that, in spite of indeed featuring overhangs and
relatively large local orientation angles, interfaces described by
Eq.\ (\ref{growth.eq}) follow a standard Family-Vicsek scaling, with
exponents that fit into the standard universality classes. For smaller
systems or low values of $A_n$, we obtain exponent values that are not
far from those of the Edwards-Wilkinson (EW) universality class in 1D,
i.e., $\alpha_{\rm EW}=\alpha_{{\rm EW},loc}=1/2$, $\beta_{\rm
  EW}=1/4$, $z_{\rm EW}=2$.  This is the behavior associated with the
weak coupling regime of the KPZ equation.  Indeed, recall that in the
Monge gauge the (naive) KPZ coupling constant scales as the square of
the noise amplitude \cite{Barabasi}. If $L$ and $A_n$ are large
enough, and within error bars, we obtain exponent values compatible
with those of the (strong coupling) 1D Kardar-Parisi-Zhang
universality class, namely, $\alpha_{\rm KPZ}=\alpha_{{\rm
    KPZ},loc}=1/2$, $\beta_{\rm KPZ}=1/3$, $z_{\rm KPZ}=3/2$.  It is
also remarkable that, in contrast to the no-overhang and small slopes
case, $A_0$ plays no role in the determination of the universality
class.

The system size of the simulations span a wide range of length scales,
from $L=0.5$ to $L=500$, always ensuring that the continuum limit has
been achieved. For the largest sizes, where KPZ scaling is found, we
have used $\Delta t=5\cdot 10^{-3}$, and spatial cutoffs
$l_{min}=5\cdot 10^{-5}L$ and $l_{max}=5\cdot 10^{-3}L$. With these
values, the number of points composing our simulated interfaces range
typically between $100 \lesssim N \lesssim 1000$.  Simulation times
have reached $t=7000$ in order to ensure saturation to a steady state.
Equation parameters have been taken of order unity: $A_0\in [0,1]$,
$A_1\in [0,1]$ and $A_n\in [0,3]$. Ensemble convergence is ensured by
taking, in most cases, 200 samples or more.

Roughness measurements are done using our diagonal-fitting algorithm,
although checks have been performed with the usual Monge-gauge
definition, with no change in the exponents. We should remark that the
simulation algorithm is extremely robust and does not suffer from
instabilities.
%Moreover, the scaling behaviour is always very clean,
%as it can be seen from the plots.

\begin{figure}[t]
\epsfig{file=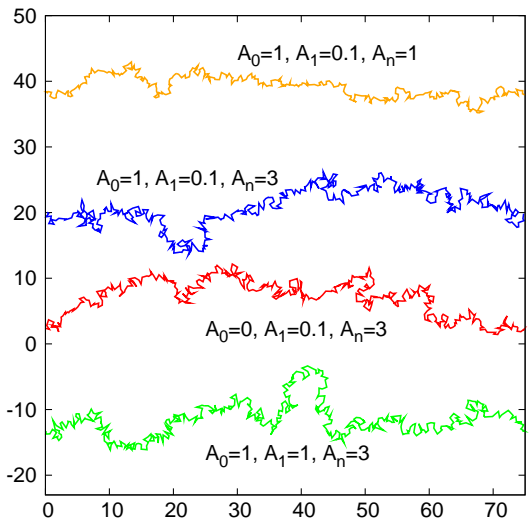,width=7cm,angle=270}
\caption{\label{fig.profiles} Typical interface profiles for $L=75$ at
  saturation ($t=3000$) for different values of the parameters as indicated.
  The curves have been artificially offset in the vertical direction in order
  to allow comparison. All units are arbitrary.}
\end{figure}

\begin{figure}[t]
\epsfig{file=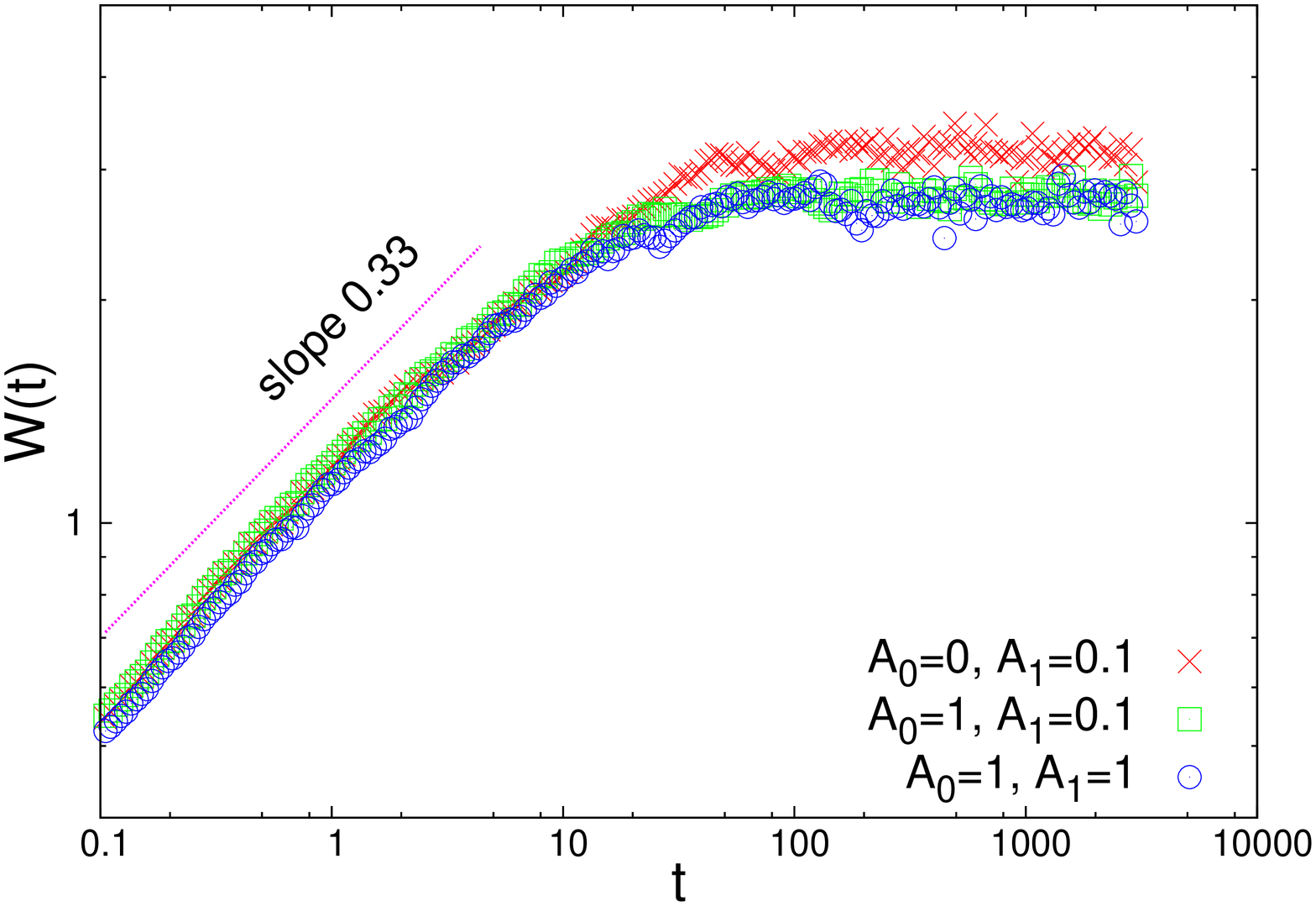,width=5.7cm,angle=270}
\epsfig{file=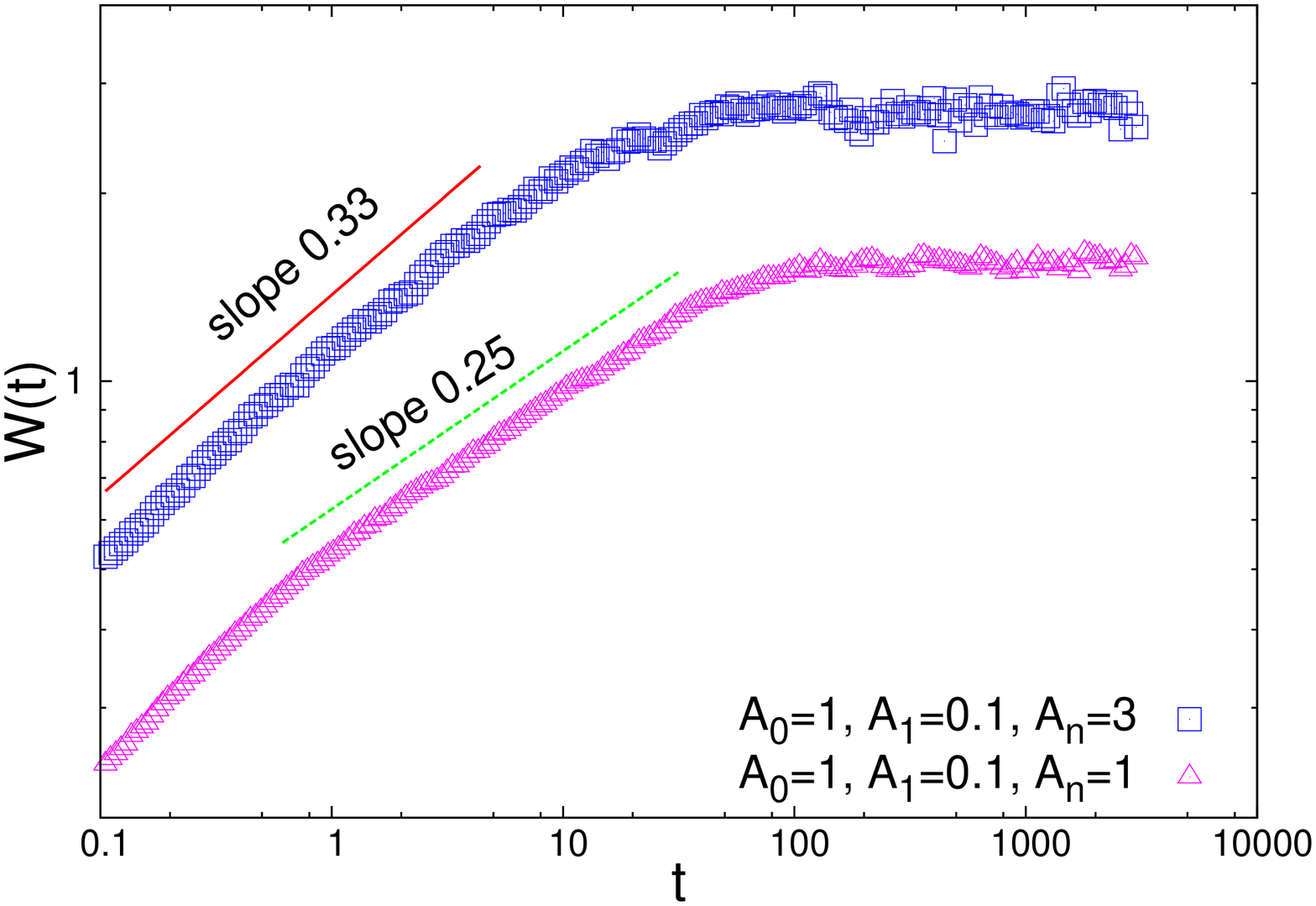,width=5.7cm,angle=270}
\caption{\label{fig.width} Time evolution of the global interface
  width for $L=75$ and different parameter choices. Left panel: for
  $A_n=3$, the scaling behavior does not depend appreciably on the
  values of $A_0$ and $A_1$. The straight line is a fit to the data
  within the corresponding time interval, its slope being close to the
  KPZ behavior $\beta_{\rm KPZ}=1/3$. Right panel: comparison between
  $A_n=3$ case shown in the left panel and $A_n=1$.  The straight
  lines are obtained from fits to the data within the corresponding
  time intervals, having slopes that are close to the KPZ and EW
  behaviors, $\beta_{\rm KPZ}=1/3$ and $\beta_{\rm EW}=1/4$,
  respectively. All units are arbitrary.}
\end{figure}

Fig.\ \ref{fig.profiles} shows three typical interface profiles for
$L=75$, for the same times ($t=3000$), once saturation is achieved,
for different values of the parameters. Note the relatively large
slope values and the presence of overhangs. As intuitively expected,
comparison between the orange and the blue curves illustrates the
increased abundance of overhangs for larger noise
intensities. However, changing from the latter to the red interface,
we are led to expecting little influence of the value of $A_0$ on the
morphological properties. In turn, the green line corresponds to a
parameter condition that differs from that of the blue line in the
value of $A_1$. As seen in Fig.\ \ref{fig.width}, this also seems to
have little effect in the scaling behavior.

The time evolution of the global interface width is shown in the left
panel of Fig.\ \ref{fig.width} from $t=0.1$ to $t=3000$ for $L=75$,
with $200$ samples and $A_n=3$. For all shown cases, prior to
saturation there is a scaling regime characterized by a exponent value
$\beta = 0.33\pm 0.01$ as given by a fit for the time range indicated
in the figure.  The overlap of the three curves is remarkable, this
scaling behavior appearing to be quite independent on parameter $A_0$
being non-zero. Notice that equation (\ref{growth.eq.complete})
corresponds to non-equilibrium (irreversible) growth even if $A_0=0$,
due to removal of self-intersections. The behavior shown in
Fig.\ \ref{fig.width} provides evidence on our claim that $A_0$ is
irrelevant to the scaling behavior. We reach a similar conclusion from
the measurements of additional exponents, although the results shown
below will mostly consider $A_0\neq 0$ conditions.

Additionally, the value of $A_1$ is seen in Fig.\ \ref{fig.width} to have little
effect in the strong coupling KPZ scaling behavior.
However, for smaller system sizes or smaller noise amplitudes, the
observed value of the growth exponent $\beta\simeq 0.25$ is closer to
that of the Edwards-Wilkinson universality class, see the right panel
in Fig.\ \ref{fig.width}, suggesting the system is in a preasymptotic
transient for such parameter conditions, as expected from results in
the Monge gauge \cite{Katzav_PRE04}, compare both panels of our
Fig.\ \ref{fig.width} with e.g.\ Figs.\ 2 and 4 in
\cite{Giada_PRE02,Giada_PRE02b}, where accurate numerical simulations
of the standard KPZ equation are reported using a pseudospectral
method.

\begin{figure}[t]
\epsfig{file=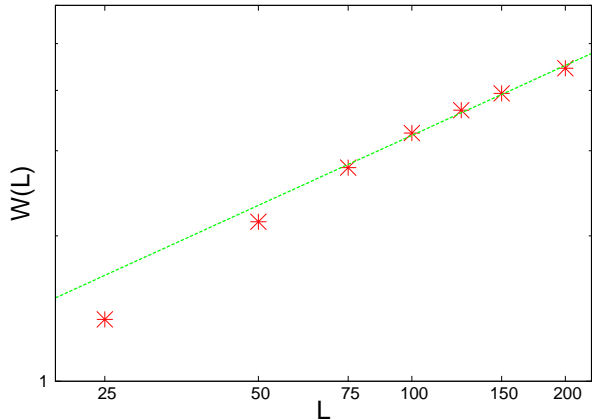,width=5.7cm,angle=270}
\caption{\label{fig.alphaglobal} Global interface width as a function
  of system size. For comparison, the straight line is obtained from a
  fit of the behavior for $75 \leq L \leq 200$ and has a slope
  corresponding to $\alpha=0.51$, compatible with $\alpha_{\rm KPZ} =
  1/2$. In all cases $A_0=1$, $A_1=0.1$, and $A_n=3$. All units are
  arbitrary. }
\end{figure}

Regarding the value of the roughness exponent, indeed
Fig.\ \ref{fig.alphaglobal} also shows the approach to an asymptotic
value compatible with the KPZ universality class for sufficiently
large system sizes $L$. Thus, the global width is plotted in this
figure as a function of $L$ for fixed $A_0=1$, $A_1=0.1$, and
$A_n=3$. In the figure, we estimate $\alpha=0.51 \pm 0.03$ from a fit
of the behavior for the largest system sizes $75 \leq L \leq 200$. In
general, the transition from non-KPZ preasymptotic (small $L$, $A_n$
values, weak coupling regime) to asymptotic KPZ (large $L$, $A_n$
values, strong coupling regime) scaling behavior is not particularly
correlated with an increasing dynamical role of self-intersection
removal. The simulations show that the length of interface that is
removed per unit time reaches a constant non-negligible value for
generic parameter and system size conditions.

\begin{figure}[t]
\epsfig{file=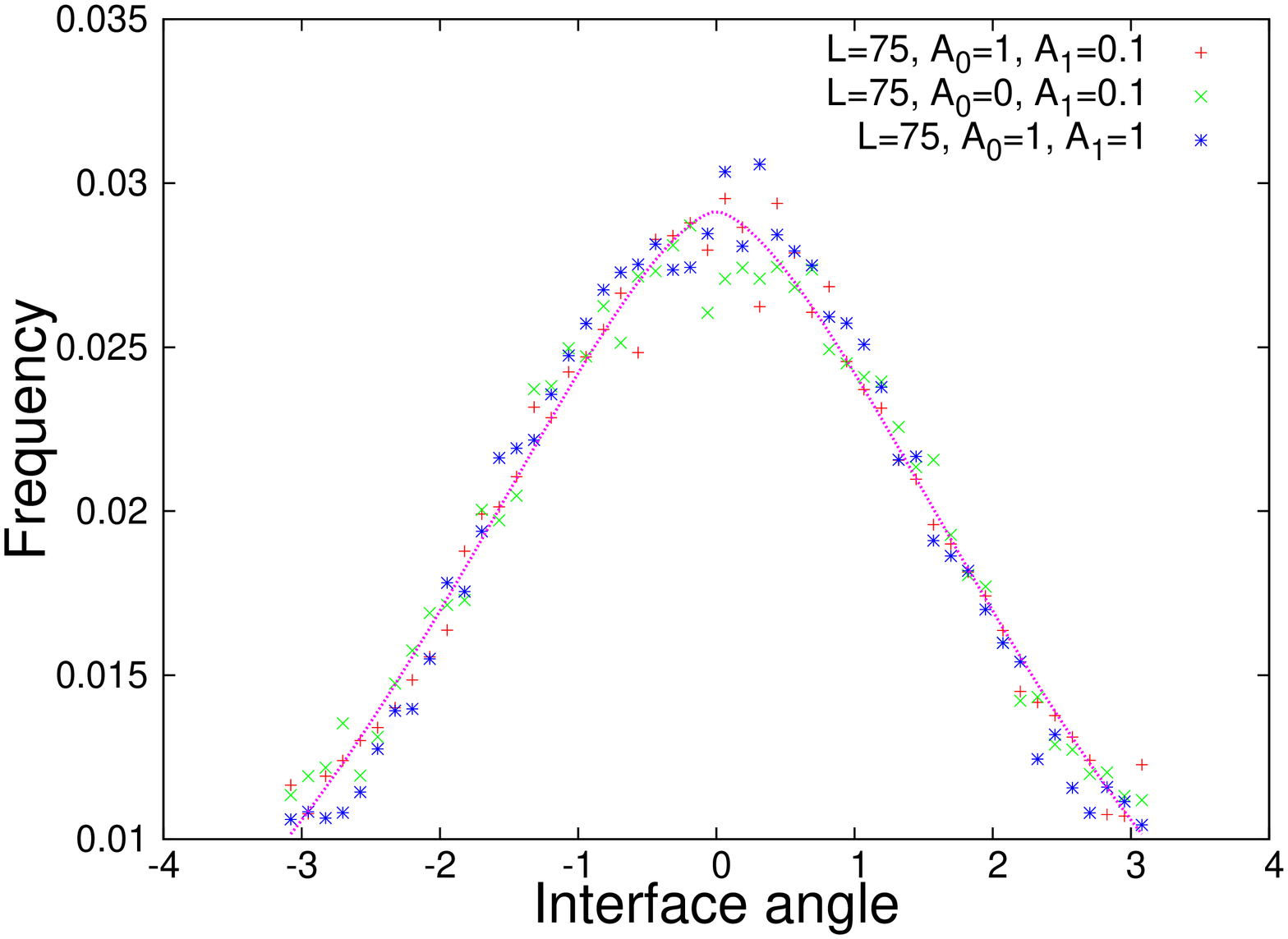,width=5.7cm,angle=270}
\epsfig{file=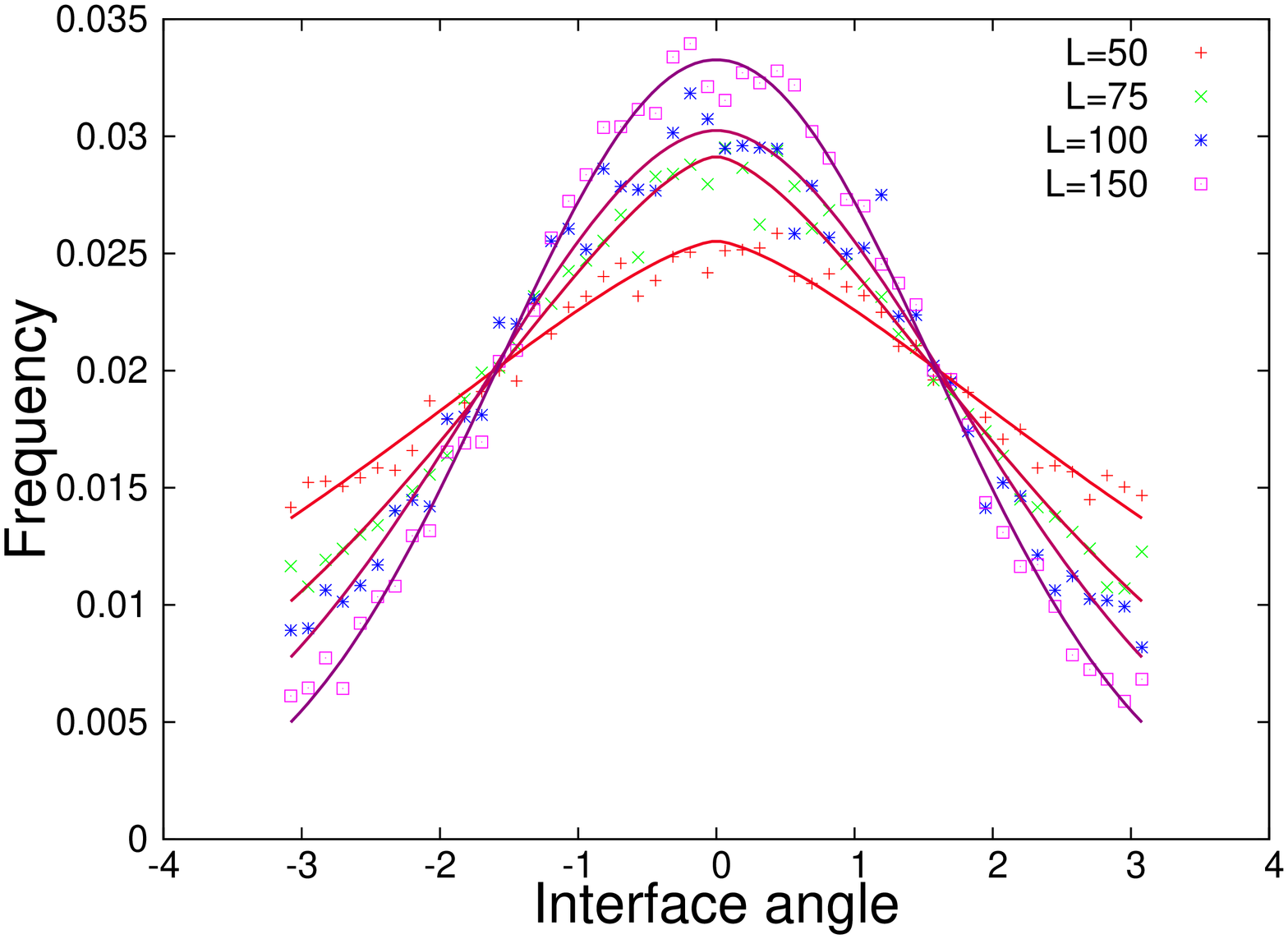,width=5.7cm,angle=270}
\caption{\label{fig.slopes} Left panel: angle histograms are shown at
  saturation for different choices of $A_0$ and $A_1$, $L=75$, and
  $A_n=3$, obtaining very similar results. The solid line is a fit to
  a shape $C_1\exp(-C_2|\theta|^\chi)$ with $\chi=1.54$. Right panel:
  for fixed $A_0=1$, $A_1=0.1$ and increasing size, the angle
  histograms can be seen to evolve towards a Gaussian profile. Solid
  lines are fits similar to that in the left panel with $\chi = 1.44,
  1.54, 1.84, 1.99$ for increasing $L$. }
\end{figure}

The large slopes and overhangs seen in Fig.\ \ref{fig.profiles}
suggest a study of the statistics of the {\em local} angle $\theta$
shown by the interface with respect to the {\em global} horizontal
direction.  The left panel of Fig.\ \ref{fig.slopes} shows the angle
histogram (i.e., angle probability distribution) for $L=75$ and
different values of the parameters, once saturation was achieved. A
fit is shown to an exponential distribution of the form
$C_1\exp(-C_2|\theta|^\chi)$. Again, $A_0$ does not seem to play any
role in this behavior, and the three curves overlap. The right panel
shows the evolution of this histogram as $L$ increases. Fits to the
same exponential form are also displayed. The corresponding values of
$\chi$ increase with $L$, stabilizing to a value close to $2$, i.e.,
the distribution becomes Gaussian.  In any case, the symmetric
exponential distribution (as opposed to a power-law one) centered
around the origin implies that the system has a {\em characteristic
  slope} that is horizontal.  Of course, the probability distribution
can not be truly Gaussian, since the angle domain is bounded in
$[-\pi,\pi]$.

\begin{figure}[t]
\epsfig{file=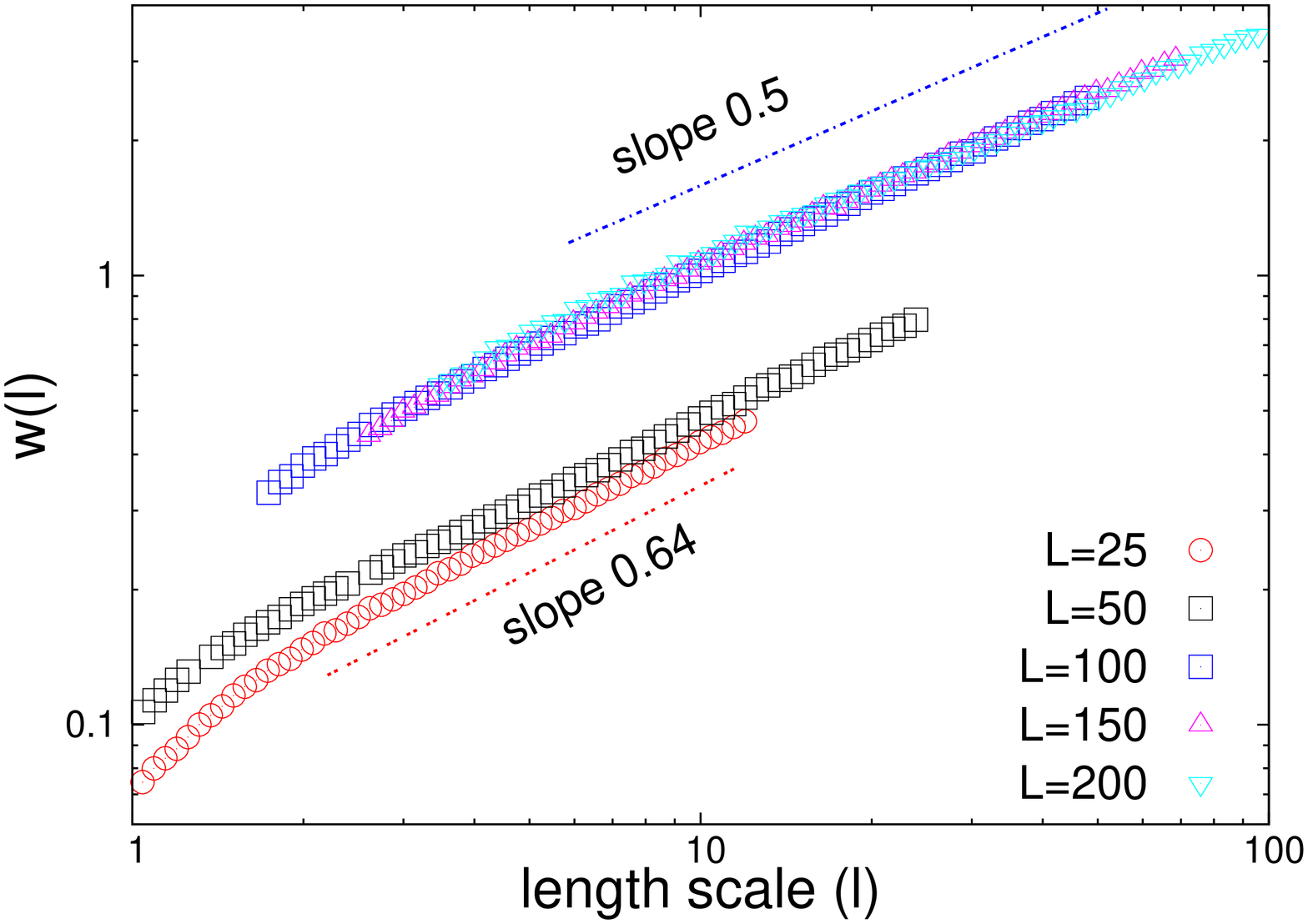,width=5.7cm,angle=270}
\epsfig{file=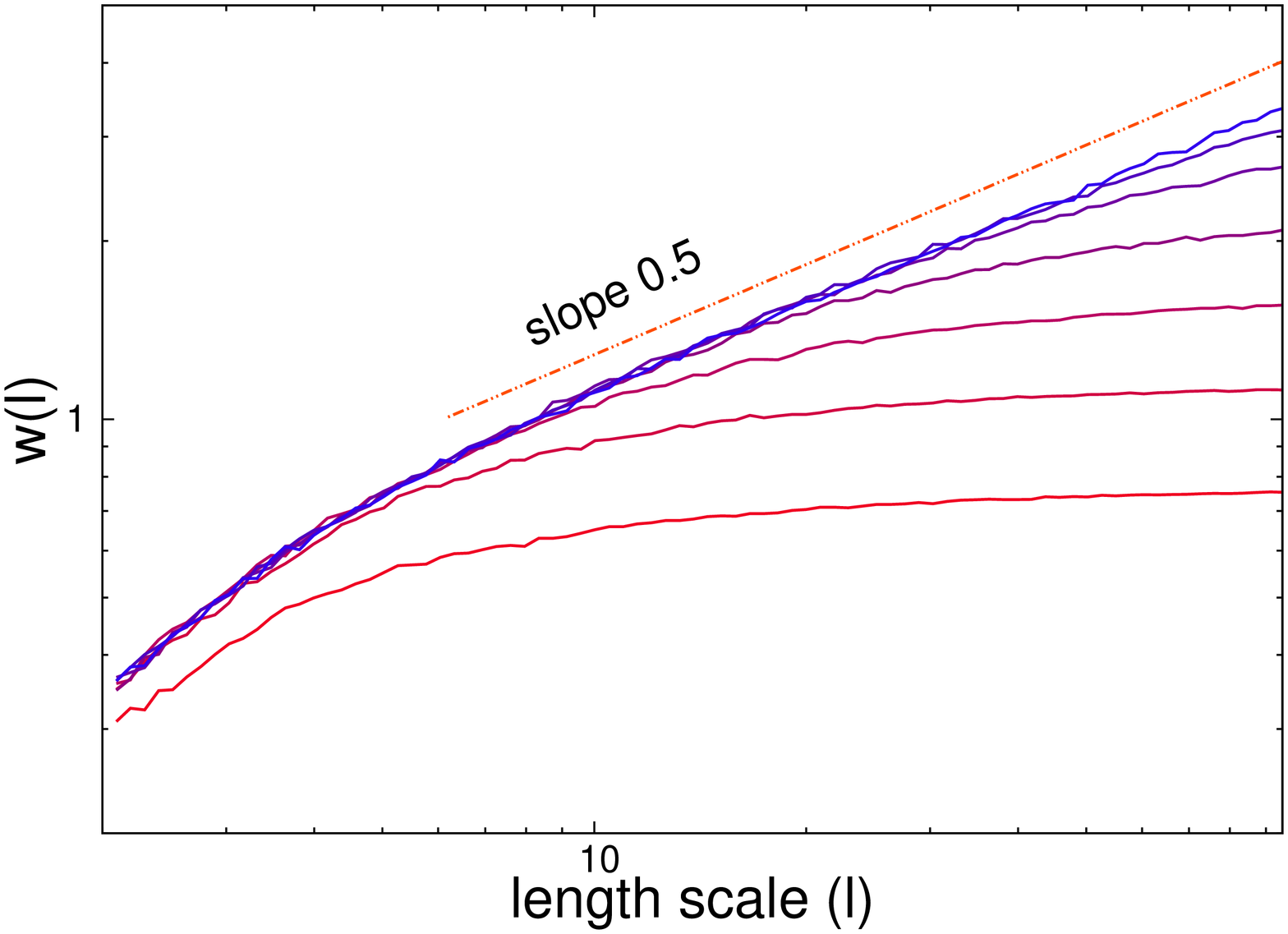,width=5.7cm,angle=270}
\caption{\label{fig.morphology} Left panel: local interface width as a
  function of scale, $w(l)$, at saturation for different system sizes
  as indicated. In all cases, $A_0=1$, $A_1=0.1$, and $A_n=3$. The two
  groups of curves ($L < 75$ and $L > 75$) have been artificially
  offset for clarity.  Notice the overlap between the different curves
  for $L \geq 100$, implying that scaling is FV.  The small $L< 100$
  behavior is slightly anomalous, and characterized by a different
  value of $\alphaloc$. The solid lines are obtained from fits for the
  smallest and largest values of $L$ in the corresponding regions of
  $l$. Their slopes provide the value $\alphaloc$ in each case. Right
  panel: local interface width $w(l,t)$ vs scale for $L=200$ and
  logarithmically equispaced times between $t=0.1$ and $t=200$, bottom
  to top. The straight line with slope $\alphaloc = 0.5$ is obtained
  from a fit of the behavior for the top curve within the
  corresponding range of $l$.  All units are arbitrary. }
\end{figure}

The change in the {\em local} morphological properties for increasing
size $L$ goes beyond the angle histogram.  Thus, the local roughness
exponent $\alphaloc$ is close to $1/2$ for small noise amplitudes
$A_n\leq 1$. If the noise amplitude is large ($A_n=3$), for small
systems ($L<50$) this exponent takes an effective value that can raise
even above $0.6$, see left panel on Fig.\ \ref{fig.morphology} where
the local interface width at saturation, $w(l)$, is shown as a
function of the local scale, for different system sizes $L$ and fixed
$A_0=1$, $A_1=0.1$, and $A_n=3$. Moreover, the non-complete overlap of
e.g.\ the curves for $L=25$, $50$, is reminiscent of anomalous scaling
behavior. However, for larger values of $L > 75$ the local roughness
exponent converges to an $L$-independent value $\alphaloc= 0.51 \pm
0.01$, namely, close within error bars to the $1/2$ value it takes
both for the standard EW and KPZ equations in 1D. Moreover, for these
system sizes curves overlap, a fact that indicates that asymptotic
scaling is not anomalous.
%This effect can be appreciated in the left panel of figure \ref{fig.morphology},
%For the data that are shown, we can see that curves for $L< 75$ are characterized by $\alphaloc \simeq
%0.59$ while curves for $L> 75$ share a smaller value $\alphaloc \simeq 0.51$. Within each system size regime,
%curves overlap, a fact that indicates that scaling is not anomalous.
The right panel in the same figure shows how the local width $w(l,t)$
curves evolves with time for our largest system size, $L=200$. The
curves give the behavior of the local width with $l$ for different
times.  Apart from times that are comparable with the discretization
unit, for which no proper correlations have built up, curves for
increasing times overlap in the form that again is expected for FV
scaling. As time evolves, the lateral correlation length $\xi(t)$
increases, finally reaching a steady state value at which $\alphaloc
\simeq 0.5$ is obtained from a fit of the large $l$ behavior,
compatible with the value of the global roughness exponent $\alpha$
found above.

\begin{figure}[t]
\epsfig{file=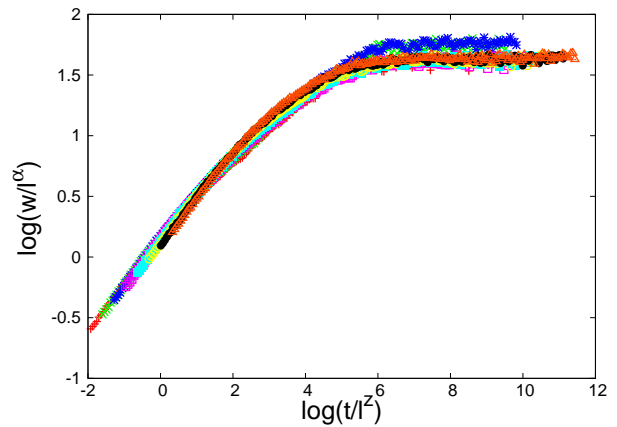,width=5.7cm,angle=270}
\caption{\label{fig.collapses} Data collapse of the local interface width $w(l,t)$ curves of the right panel on
Fig.\ \ref{fig.morphology} ($L=200$). Collapse is achieved for $\alpha=0.52$ and $z=1.4$. All units are arbitrary.}
\end{figure}

Actually, we can employ the {\em local} morphological behavior just
considered in order to measure the dynamic exponent $z$, as an
alternative check to assess the asymptotic behavior of our system. As
mentioned above, within the Family-Vicsek Ansatz, the saturation time
$t_s \sim L^z$, a relation which is also fulfilled for local scales,
$t_s(l) \sim l^z$. This implies that, if time is measured in units of
$l^z$ and the local interface width is measured in units of
$l^\alpha$, all the $w(l,t)$ curves should collapse.  This is shown in
Fig.\ \ref{fig.collapses} for the $L=200$ curves shown on the right
panel of Fig.\ \ref{fig.morphology}.  As usual \cite{Barabasi}, the
values of $\alpha$ and $z$ are chosen so that the overlap between the
curves is maximum. This occurs for $\alpha=0.52\pm 0.2$ and $z=1.4\pm
0.1$, again in agreement with the KPZ values for the roughness and the
dynamic exponents, respectively.

%{\bf PUNTO EXTRA NO INCLUIDO:} Role of the removal of
%self-intersections. Even when $A_0=0$ the system is out of
%equilibrium. The length of interface removed per unit time reaches a
%constant. This removal of self-intersections remains as the key
%mechanism for relaxation of fluctuations in the KPZ regime.

\section{Conclusions and Outlook}
\label{conclusions}

We have characterized the scaling properties of a continuum stochastic
model of one-dimensional interfaces, given by
Eq.\ (\ref{growth.eq}). The model captures the main physical
mechanisms believed to characterize non-conserved, irreversible
surface growth, as described in particular within the KPZ approach to
non-conserved interface dynamics, namely, growth along the local
normal direction with a rate that is affected by local curvature and
by fluctuations in growth events, but without assumptions on small
slopes or absence of overhangs. Our numerical simulations are tailored
to avoid these simplifications, and lead to a number of results. In
spite of the possibility for the surface to develop arbitrarily large
slopes, a fact that is frequently associated with anomalous scaling,
the kinetic roughening properties fulfill, rather, the Family-Vicsek
Ansatz. There are predictions on the non-occurrence of non-trivial
anomalous scaling at the steady state of local growth equations
\cite{Lopez_PRL05}. In principle, these predictions are done within a
small slope approximation so that our present result supports their
validity beyond such constraint for the specific example that we have
studied. This suggests in turn the interest of assessing the
corresponding situation for other continuum models beyond the small
slope approximation.

Regarding the specific type of scaling that we find, for system sizes
and/or noise amplitudes that are small, the system is in a
preasymptotic state within which the growth exponent is close to that
of the EW universality class, while the roughness exponent is close to
but not necessarily equals the corresponding EW value. In any case,
for sufficiently large system sizes and/or noise amplitudes,
Eq.\ (\ref{growth.eq}) turns out to be in the KPZ universality class.
Naturally, this result agrees with general considerations based on
symmetries and conservation laws \cite{Barabasi,Krug_AP97}, but we
find it to be non-trivial. To the best of our knowledge, it is found
for the first time for a SPDE that can access height configurations
that are not constrained by the small slope and no-overhang
approximations. Note moreover that, as a result of the very
formulation of Eq.\ \eqref{growth.eq}, the role of noise in this
system is very far from the standard KPZ case. Thus, in the latter
noise is decorrelated only in the $x$-axis. Adoption of that type of
noise in our system would amount to assuming that all interface points
which share the same $x$-coordinate are correlated, disregarding their
distance in the vertical direction. In our case, noise is decorrelated
both in the horizontal and vertical directions, as seen in
Eq.\ \eqref{noise}. Thus, its naive spatial scaling dimension should
be 2 rather than 1. Even a power counting argument \cite{Barabasi} on
the KPZ equation with this modification leads to a different result,
since it provides consistent scaling exponents which differ from the
KPZ universality class.  Thus, our present numerical results are not
evident from this point of view.

In our simulations, we have seen moreover that our extra condition on
self-intersection removal plays a non-negligible role, in particular
driving the system out of equilibrium even when $A_0=0$.  Note that
the equilibrium problem associated with Eq.\ (\ref{growth.eq}) is well
understood \cite{Plischke}. In our case, intersection removal is the
price we have to pay in order to have mathematically a
simply-connected interface. Physically, it might be thought of as
playing a similar simplifying role as the condition of a phantom
membrane plays with respect to more realistic polymerized membranes
\cite{Plischke}. Still, we believe the ensuing model features
non-trivial physics. Perhaps this can be appreciated better if we
think of still more realistic models of growth, of which
Eq.\ (\ref{growth.eq}) and similar equations are simplified
versions. In those, the dynamics of the interface is coupled to that
of a physical field, such as e.g.\ the concentration of aggregating
units in the dilute phase, in what mathematically is described as a
(stochastic) moving boundary problem, see
e.g.\ \cite{Saito,Nicoli_JSTAT09}. For parameter conditions in which
kinetic roughening occurs, voids and bubbles can indeed be created
(leading to a multiply connected interface), see
e.g.\ \cite{Keblinski_PRE96}, but their role for the morphological
evolution seems to be quite marginal
\cite{Nicoli_JSTAT09}. Incidentally, self-intersection removal seems
to be a relevant stabilizing mechanism in face of morphological
instabilities, as in ox-bow formation in meandering rivers
\cite{Edwards_PRE02,Liverpool_PRL95}. In stable-growth moving boundary
problems, the main role in the evolution is played, rather, by the
envelope of the so-called active growth zone, that corresponds in our
model to the simply connected interface that we keep track of after
self-intersection removal. From this point of view, our choice does
not seem too distant from the standard procedure by which the full
interface dynamics of e.g.\ discrete growth models that lead to
bubbles and overhangs due to e.g.\ bulk vacancies
\cite{Schimschak_PRB95} is traded for that of its single valued
approximation.

An interesting future extension of the present work is the study of
the full Eq.\ \eqref{growth.eq.complete}, namely, for a non-zero $A_2$
coefficient, that would allow us to study surface diffusion within the
present framework, with potential implications in the context of thin
film growth by MBE \cite{Pimpinelli}. Given the slow dynamics associated
with relaxation by surface diffusion, this would require improvements on
the more computationally expensive ingredients of our adaptive numerical
scheme, conspicuously including the self-intersection removal algorithm.

Moreover, given the geometrically intrinsic nature of the dynamics,
the numerical approach, and the measurement techniques developed in
the present work ---by which nonlinearities and noise correlations,
which would appear in the Monge gauge, come out naturally---,
additional interesting applications would be the study of growth in a
circular geometry (see a recent discussion in \cite{Escudero_JSTAT09}
and references therein) and the extension to two-dimensional
interfaces. In both cases there are still interesting open theoretical
and experimental challenges.

\begin{acknowledgments}
We would like to thank the two anonymous reviewers for their insightful
remarks and recommendations, as well as Carlos Escudero
for useful discussions. S.\ N.\ S.\ thanks the plasma physics group of 
Universidad Carlos III de Madrid for granting her access to their computational 
facilities. This work has been supported by Spanish MICINN
grant No.\ FIS2009-12964-C05-01. 
\end{acknowledgments}

%\section*{References}
% Comment if you want to insert a .bbl file directly below
\bibliography{surf}

\end{document}